\documentclass{article}
\usepackage{spconf,amsmath,graphicx}
\usepackage{verbatim}
\usepackage{wrapfig}

\graphicspath{{./}}
\usepackage{graphics,graphicx,url,color}
\usepackage{amsmath,amsfonts,amssymb,theorem,euscript,mathrsfs,bbm}
\newtheorem{theorem*}{Theorem}
\usepackage{epsf,epsfig,wrapfig}
\usepackage{array,enumitem,multicol}
\usepackage{footmisc}
\DeclareGraphicsExtensions{.ps}

\definecolor{navyblue}{rgb}{0.0, 0.0, 0.5}
\definecolor{lightblue}{rgb}{0.3, 0.3, 0.9}

\newcommand{\bracketprod}[2]{\left[{\kern-.75ex}\left[#1,#2\right]{\kern-.75ex}\right]_{}}

\DeclareMathOperator*{\argmin}{arg\,min}

\newcommand{\shah}{{\textstyle \amalg{\kern-4.pt\amalg}}}

\title{Gram Filtering and Sinogram Interpolation for Pixel-basis in Parallel-beam X-ray CT Reconstruction}
\name{Ziyu Shu, Alireza Entezari}
\address{CISE Department, University of Florida, Gainesville, FL 32611-6120, USA}
\begin{document}
%
\maketitle
\begin{abstract}
The key aspect of parallel-beam X-ray CT is forward and back projection, but its computational burden continues to be an obstacle for applications. We propose a method to improve the performance of related algorithms by calculating the Gram filter exactly and interpolating the sinogram signal optimally. In addition, the detector blur effect can be included in our model efficiently. The improvements in speed and quality for back projection and iterative reconstruction are shown in our experiments on both analytical phantoms and real CT images.
\end{abstract}
\section{Introduction}
\label{sec:intro}
Parallel-beam CT is widely used in areas such as phase-contrast X-ray imaging \cite{doi:10.1063/1.364374}, \cite{pub.1005348937}, electron tomography \cite{MCINTOSH200543}, and single-particle cryo-electron microscopy \cite{10.1093/jmicro/dfv366}. The tomographic reconstruction problem often involves a forward model $H$ and corresponding normal operator $H^TH$. Generally, the normal operator can be expressed as a large Gram matrix, which is computationally expensive. The Parallel-beam CT has the property that its normal operator is linear and shift-invariant under band limited assumption. The result is that we can replace the large Gram matrix by a relatively small Gram filter to accelerate iterative reconstruction algorithm. This was discovered by \cite{495957} and recently extended to 3D by \cite{McCann:16}. State-of-the-art methods such as \cite{McCann:16} and \cite{7932483} use separable sinc function as discretization kernel to utilize this property. However, due to its space unlimited property, sinc function is not a practical kernel for application, its estimation is made at the cost of accuracy.

In this paper, we provide a formulation for the efficient Gram filtering of X-ray transform and optimized sinogram signal interpolation in parallel-beam geometry for the commonly-used pixel-basis instead of the sinc basis. This formulation removes the band limited assumption and yet improves the speed and accuracy in the reconstruction process. Furthermore, we can model the detector blur effect efficiently. Quality of back projection and reconstruction is improved significantly with such basis.

The paper is organized as follows: In Section 2, we formulate the 2D parallel-beam X-ray reconstruction problem. In Section 3, we introduce the box spline, which coincides with pixel-basis, and its X-ray transformation property. In Section 4, we explain how box spline helps us calculate the Gram matrix. In Section 5, we explain how box spline makes interpolation more accurate during the back projection. The proposed method is compared with standard methods in terms of speed, SNR and SSIM, for back projection and reconstruction of phantoms and real CT images.

\section{Problem Formulation}
\label{sec:problem setting}
The attenuation map of the imaged object $f_c$ is discretized using a discretization kernel $\psi$ as $f_{\psi}(\boldsymbol{x}) = \sum_{\boldsymbol{k}} c_{\boldsymbol{k}}\psi(\boldsymbol{x} - \boldsymbol{\Lambda}_x \boldsymbol{k})$, where $c_{\boldsymbol{k}}$ is the set of coefficients of total number $N^2$, $\boldsymbol{\Lambda}_x$ is a diagonal matrix specifying the sampling step in each dimension. The X-ray transformation operator $\mathcal{P}_{\boldsymbol{\theta}}$ specified by $\boldsymbol{\theta}$ transforms $f_c$ to: $g_c^{\boldsymbol{\theta}}  = \mathcal{P}_{\boldsymbol{\theta}} f_c$ and over all angles as $g_c = \mathcal{P} f_c$. The sampled data, observed in the sinogram domain, is $g_s^{\boldsymbol{\theta}} = S g_c^{\boldsymbol{\theta}}$, where $S$ is the sampling operator defined as $S g_c^{\boldsymbol{\theta}}[m] = \langle \delta(y-\lambda_y m), g_c^{\boldsymbol{\theta}}(y) \rangle,$ where $m \in \mathbb{Z}$, $\delta$ is Dirac's $\delta$ function, $\lambda_y$ is the sampling step in sinogram domain, $\langle a, b \rangle$ indicates the inner product between $a$ and $b$.

Without detector blur effect, the forward model is just the X-ray transformation, $H = \mathcal{P}$. In practice, due to the finite extent of a detector cell, a blurring effect is introduced in the forward model. The kernel function of the detector cell can be modeled by convoluting a series of box functions with different heights and lengths. For simplicity, only one box function is used in this paper, so that the integration over the detector cell can be modeled as a sampling of the sinogram signal convolved with the box function which is scaled according to the detector cell width $\zeta_{\rm blur}$. Therefore, the forward model becomes to $H = \frac{1}{\zeta_{\rm blur}}{\rm box}(\frac{\cdot}{\zeta_{\rm blur}})*\mathcal{P}_{\boldsymbol{\theta}}$.

Our goal is to find $\hat{f}_\psi$ to minimize the total error $E_{\rm total}$:
\begin{equation}\nonumber
E_{\rm total} = f_c - \hat{f}_\psi = (f_c - f_\psi) + (f_\psi - \hat{f}_\psi) = E_{d} + E_{r},
\end{equation}
where $f_\psi$ is the best representation of $f_c$ with discretization kernel $\psi$, $\hat{f}_\psi$ is the optimized result from reconstruction algorithms. $E_d$ is the discretization error only depends on the selection of discretization kernel. $E_d$ vanishes as the sampling step tends to zero when discretization kernel satisfied the partition-of-unity property \cite{843002}. In the X-ray reconstruction, we have $g_s$ and aim to minimize the reconstruction error $E_r$ by solving the optimization problem:
\begin{equation}\nonumber
\argmin_{\hat{f}_\psi} ||\hat{g}_\psi - H\hat{f}_\psi||^2,    
\end{equation}
where $\hat{g}_\psi$ is the estimation of $g_\psi = Hf_\psi$ by interpolating $g_s$. Current applications in this field make band limited assumption, then the $N^2 \times N^2$ Gram matrix can be represented by a $(2N - 1) \times (2N -1)$ Gram filter. In this case, we can execute iterative reconstruction with a single back projection, $H^T \hat{g}_{\psi}$, at the beginning and apply the filter in every iteration. 

The main drawback of the band limited assumption is the loss of accuracy. To calculate the filter, truncation is made due to the discretization of separable sinc function \cite{7932483}. Also, prefiltering, truncation or oblique interpolation is used for back projection since sinc function is used to interpolate the sampled data given in $g_s$ \cite{495957}, \cite{McCann:16}, \cite{7932483}. To solve these problems, we use pixel-basis (bivariate box spline) as discretization kernel instead.

\section{Image Discretization with Box Splines}

\begin{wrapfigure}{r}{0.26\textwidth}
	\centering
	\vspace{-10pt}
	\includegraphics[width=0.2\textwidth]{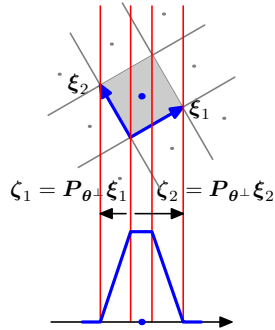}
	\vspace{-10pt}
	\caption{X-ray transform of pixel-basis as univariate box spline.}
	\label{pixelbasis}
\end{wrapfigure}

A box spline is a function ($\mathbb{R}^d \rightarrow \mathbb{R}$) defined by a set of $N$ vectors $ \boldsymbol{\Xi} = [ \boldsymbol{\xi}_1, \boldsymbol{\xi}_2, ..., \boldsymbol{\xi}_N ]$ in $\mathbb{R}^d$. An elementary box spline, $M_{\boldsymbol{\xi}}$, is a Dirac-like generalized function supported on $t\boldsymbol{\xi}$ for $0 \leq t \le 1$. The general box spline is defined as the convolution of distributions associated with the single-vector box splines:

\begin{equation} \label{eq:boxconv}
    M_{\boldsymbol{\Xi}}(\boldsymbol{x}) = M_{\boldsymbol{\xi}_1} * M_{\boldsymbol{\xi}_2} * ... * M_{\boldsymbol{\xi}_N}(\boldsymbol{x}). 
\end{equation}

The X-ray transformation of a bivariate box spline $M_{\boldsymbol{\Xi}}$ is a univariate box spline whose direction set $ [\boldsymbol{\zeta}_1, \boldsymbol{\zeta}_2, ...,\boldsymbol{\zeta}_N]$ is the geometric projection of the direction set $\boldsymbol{\Xi}$ \cite{6172241}, specifically:

\begin{equation} \label{eq:boxxray}
    \mathcal{P}_{\boldsymbol{\theta}} M_{\boldsymbol{\Xi}}({y}) = M_{\boldsymbol{P}_{\boldsymbol{\theta}^\perp} \boldsymbol{\Xi}}({y}),
\end{equation}
where $\boldsymbol{P_{\theta^\perp}}$ is the transformation matrix that geometrically projects the $\boldsymbol{x}$-coordinate system onto the ${y}$-coordinate system perpendicular to $\boldsymbol{\theta}$.

As shown in Fig.\ref{pixelbasis}, pixel-basis coincides with two directions bivariate box spline, its X-ray transformation coincides with two directions univariate box spline. Also, the scaled box function we used to model the detector blur effect can be represented by a univariate box spline, $\frac{1}{\zeta_{\rm blur}}\rm box (\frac{\cdot}{\zeta_{\rm blur}}) = \frac{1}{\zeta_{\rm blur}}M_{\boldsymbol{\zeta}_{\rm blur}}$.

\section{Gram matrix}
By using bivariate box spline, an attenuation map $f_c$ can be represented as $f_\psi(\boldsymbol{x}) = \sum_{\boldsymbol{k}} c_{\boldsymbol{k}}M_{\boldsymbol{\Xi}}(\boldsymbol{x}-\boldsymbol{\Lambda}_x\boldsymbol{k})$, where $M_{\boldsymbol{\Xi}}$ is a two directions bivariate box spline. Assume $\boldsymbol{\Xi} = [(1,0),(0,1)]$ and $\boldsymbol{\Lambda}_x$ is an identity matrix, not considering the detector blur effect, $g_\psi = Hf_\psi$ in $\boldsymbol{\theta}$ is:
\begin{equation}\nonumber
g_\psi(y) = \mathcal{P}_{\boldsymbol{\theta}} f_\psi({y}) = \sum_{{k_{\boldsymbol{\theta}}}} c_{{k_{\boldsymbol{\theta}}}} \mathcal{P}_{\boldsymbol{\theta}} M_{\boldsymbol{\Xi}}({y-k_{\boldsymbol{\theta}}}),
\end{equation}
where ${k_{\boldsymbol{\theta}}} = \boldsymbol{P_{\theta^\perp} k}.$ By using Equation \ref{eq:boxxray}, we get:

\begin{equation} \label{eq:xray_of_f}
    g_\psi(y) = \sum_{{k_{\boldsymbol{\theta}}}} c_{{k_{\boldsymbol{\theta}}}} M_{\boldsymbol{P_{\theta^\perp} \Xi}}({y-k_{\boldsymbol{\theta}}}),
\end{equation}
where $M_{\boldsymbol{P_{\theta^\perp} \Xi}}$ is a two directions univariate box spline specified by $\boldsymbol{P_{\theta^\perp} \Xi} = [\cos \theta, \sin \theta]$.

Then, the forward model $H$ can be expressed as a matrix:

\begin{equation} \nonumber
    \boldsymbol{H} = 
    \left[
    \begin{matrix}
    M_{\boldsymbol{P}_{\boldsymbol{\theta}_1^\perp}\boldsymbol{\Xi}}({y}-{k}_{1\boldsymbol{\theta}_1}) & \cdots & M_{\boldsymbol{P}_{\boldsymbol{\theta}_1^\perp}\boldsymbol{\Xi}}({y}-{k}_{N^2\boldsymbol{\theta}_1})\\
    \vdots &  & \vdots \\
    M_{\boldsymbol{P}_{\boldsymbol{\theta}_n^\perp}\boldsymbol{\Xi}}({y}-{k}_{1\boldsymbol{\theta}_n}) & \cdots & M_{\boldsymbol{P}_{\boldsymbol{\theta}_n^\perp}\boldsymbol{\Xi}}({y}-{k}_{N^2\boldsymbol{\theta}_n})
    \end{matrix}
    \right],
\end{equation}
where $n$ is the number of projection angles.

Let $\boldsymbol{G}$ denote the Gram matrix of $\boldsymbol{H}$, $\boldsymbol{G} = \boldsymbol{H}^T\boldsymbol{H}$, then the element in its $i$th row and $j$th column, $g_{ij}$, can be expressed as:
\begin{equation}\nonumber
    g_{ij} = \sum_{\boldsymbol{\theta}} M_{\boldsymbol{P}_{\boldsymbol{\theta}^\perp} \boldsymbol{\Xi} \cup \boldsymbol{P}_{\boldsymbol{\theta}^\perp} \boldsymbol{\Xi}}({k}_{i\boldsymbol{\theta}} - {k}_{j\boldsymbol{\theta}}),
\end{equation}
where $M_{\boldsymbol{P}_{\boldsymbol{\theta}^\perp} \boldsymbol{\Xi} \cup \boldsymbol{P}_{\boldsymbol{\theta}^\perp} \boldsymbol{\Xi}}$ is a four directions univariate box spline specified by the direction set $[\cos \theta, \cos \theta, \sin \theta, \sin \theta].$

Considering the detector blur effect, as mentioned in Section \ref{sec:problem setting}, the forward model becomes $H = \frac{1}{\zeta_{\rm blur}} M_{\boldsymbol{\zeta}_{\rm blur}}*\mathcal{P}_{\boldsymbol{\theta}}$. Utilizing Equation \ref{eq:boxconv}, $g_\psi$ can be expressed as:

\begin{equation} \label{eq:blurxray_of_f}
g_\psi(y) =\frac{1}{\zeta_{\rm blur}} \sum_{{k_{\boldsymbol{\theta}}}} c_{{k_{\boldsymbol{\theta}}}} M_{\boldsymbol{P_{\theta^\perp} \Xi}\cup \boldsymbol{\zeta}_{\rm blur}}({y-k_{\boldsymbol{\theta}}}),
\end{equation}
where ${\boldsymbol{P}_{\boldsymbol{\theta}^\perp}\boldsymbol{\Xi}\cup\boldsymbol{\zeta}_{\rm blur}} = [\cos\theta, \sin\theta, \zeta_{\rm blur}]$. Therefore, the forward model can still be expressed as a matrix:
\begin{multline}\nonumber
\boldsymbol{H} =\\
\frac{1}{\zeta_{\rm blur}}
\left[
\begin{matrix}
\begin{smallmatrix}
M_{\boldsymbol{P}_{\boldsymbol{\theta}_1^\perp}\boldsymbol{\Xi}\cup\boldsymbol{\zeta}_{\rm blur}}({y}-{k}_{1\boldsymbol{\theta}_1}) & \cdots & M_{\boldsymbol{P}_{\boldsymbol{\theta}_1^\perp}\boldsymbol{\Xi}\cup\boldsymbol{\zeta}_{\rm blur}}({y}-{k}_{N^2\boldsymbol{\theta}_1})\\
\vdots &  & \vdots \\
M_{\boldsymbol{P}_{\boldsymbol{\theta}_n^\perp}\boldsymbol{\Xi}\cup\boldsymbol{\zeta}_{\rm blur}}({y}-{k}_{1\boldsymbol{\theta}_n}) & \cdots & M_{\boldsymbol{P}_{\boldsymbol{\theta}_n^\perp}\boldsymbol{\Xi}\cup\boldsymbol{\zeta}_{\rm blur}}({y}-{k}_{N^2\boldsymbol{\theta}_n})
\end{smallmatrix}
\end{matrix}
\right].
\end{multline}

 We can also get its corresponding Gram matrix, whose element $g_{ij}$ can be expressed as:
\begin{equation}\nonumber
g_{ij} = (\frac{1}{\zeta_{\rm blur}})^2\sum_{\boldsymbol{\theta}} M_{\boldsymbol{P}_{\boldsymbol{\theta}^\perp} \boldsymbol{\Xi} \cup \boldsymbol{\zeta}_{\rm blur} \cup \boldsymbol{P}_{\boldsymbol{\theta}^\perp} \boldsymbol{\Xi} \cup \boldsymbol{\zeta}_{\rm blur}}({k}_{i\boldsymbol{\theta}} - {k}_{j\boldsymbol{\theta}}),
\end{equation}
where $M_{\boldsymbol{P}_{\boldsymbol{\theta}^\perp} \boldsymbol{\Xi} \cup \boldsymbol{\zeta}_{\rm blur} \cup \boldsymbol{P}_{\boldsymbol{\theta}^\perp} \boldsymbol{\Xi} \cup \boldsymbol{\zeta}_{\rm blur}}$ is a six directions univariate box spline specified by the direction set:
\begin{equation}\nonumber
[\cos \theta, \cos \theta, \sin \theta, \sin \theta, \zeta_{\rm blur}, \zeta_{\rm blur}].
\end{equation}

Note that the $N^2 \times N^2$ Gram matrix $\boldsymbol{G}$ is a Toeplitz-block-Toeplitz matrix, which implies that we can compute $\boldsymbol{Gx}$ exactly by applying a $(2N - 1) \times (2N - 1)$ filter on $\boldsymbol{x}$.

\section{Back Projection}
\label{sec:back projection}
In this section, we aim to get $H^T \hat{g}_{\psi}$ accurately. From Section 4 we can calculate $H^T$ exactly, thus what we need to do now is to optimize $\hat{g}_{\psi}$, which is the estimation of $g_{\psi}$. Note that we know the continuity of $g_\psi$, this observation provides an advantage for utilizing the Strang and Fix condition \cite{Strang2011} to improve the estimation in our method. 

From Equation \ref{eq:xray_of_f}, we know that $g_{\psi}$ is composed of a series of shifted two directions univariate box splines not considering the detector blur effect. Therefore it must be $C^0$ continuous. In that case, the signal can be recovered exactly by using B-spline of degree $1$ as interpolation kernel when all the $k_{\boldsymbol{\theta}}$ are included in the sampling points. In the case of axis aligned projection, the signal is $C^{-1}$ continuous, and B-spline of degree $0$ is used for interpolation. By using the oblique projection \cite{843002} \cite{330352} the recovered signal $\hat{g}_\psi$ can be expressed as:
\begin{equation}\nonumber
    \hat{g}_{\psi}(y) = \sum_m (g_s*q)[m]B(y-m),
\end{equation}
where $B$ indicates B-spline function, $q[m]$ is the digital correction filter \cite{843002} \cite{330352}, whose Z-transform $Q[z]$ is:
\begin{equation} \label{eq:correction filter}
    Q[z] = \frac{1}{\sum_{m \in \mathbb{Z}} b[m]z^{-m}},
\end{equation}
where $b[m] = \langle \delta(y - m), B(y) \rangle$ is the cross-correlation sequence between $\delta(y)$ and $B(y)$. For B-spline of degree $0$ or $1$, $q[m] = \delta[m]$.

From Equation \ref{eq:blurxray_of_f}, we know that $g_\psi$ is a $C^0$ or $C^1$ continuous signal considering the detector blur effect. Therefore, B-spline of degree $1$ or $2$ is used to recover the signal. From the Equation \ref{eq:correction filter}, for B-spline of degree $2$, $q[m] = (2\sqrt{2}-3)^{|m|}$ \cite{193221}. The sequence converges to zero rapidly (i.e., $q[40] = 3.38 \times 10^{-31}$) and it can be truncated without noticeable effects.

\section{Experiments and Results}
\label{sec:typestyle}
\begin{figure}[htb]

\begin{minipage}[b]{.30\linewidth}
  \centering
  \centerline{\includegraphics[width=2.5cm]{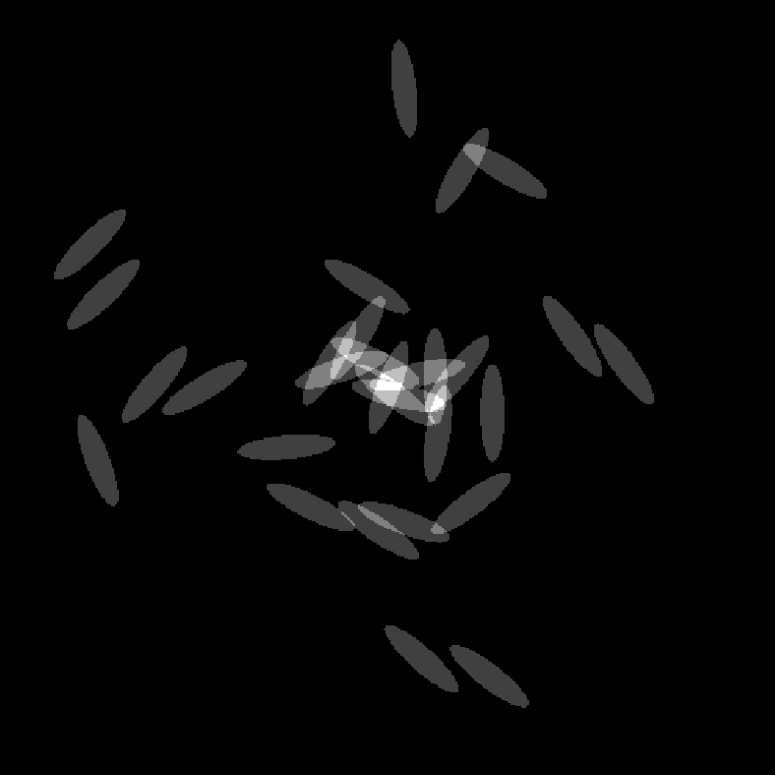}}
  \centerline{(a)}\medskip
\end{minipage}
\hfill
\begin{minipage}[b]{.30\linewidth}
  \centering
  \centerline{\includegraphics[width=2.5cm]{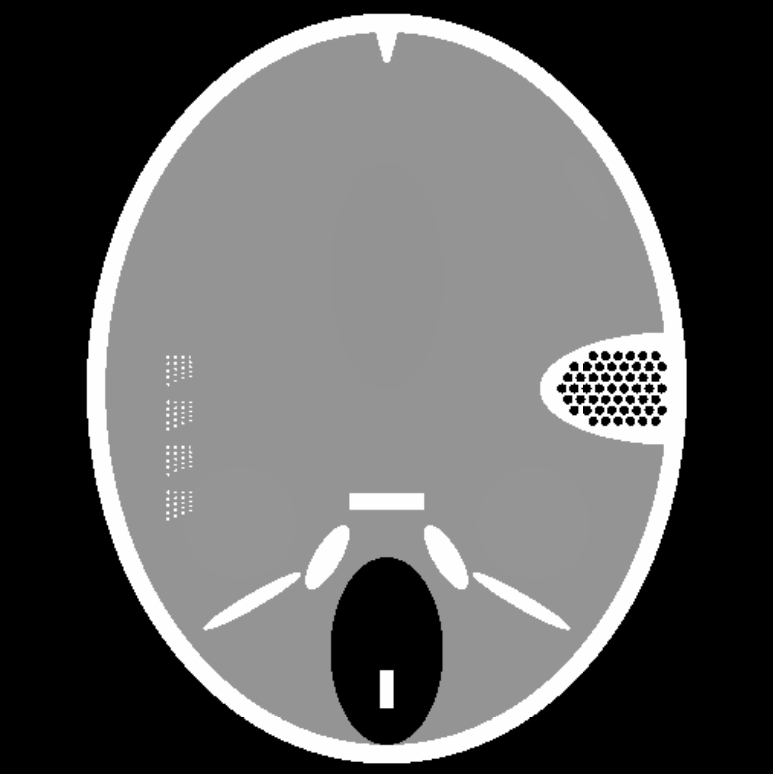}}
  \centerline{(b)}\medskip
\end{minipage}
\hfill
\begin{minipage}[b]{0.30\linewidth}
  \centering
  \centerline{\includegraphics[width=2.5cm]{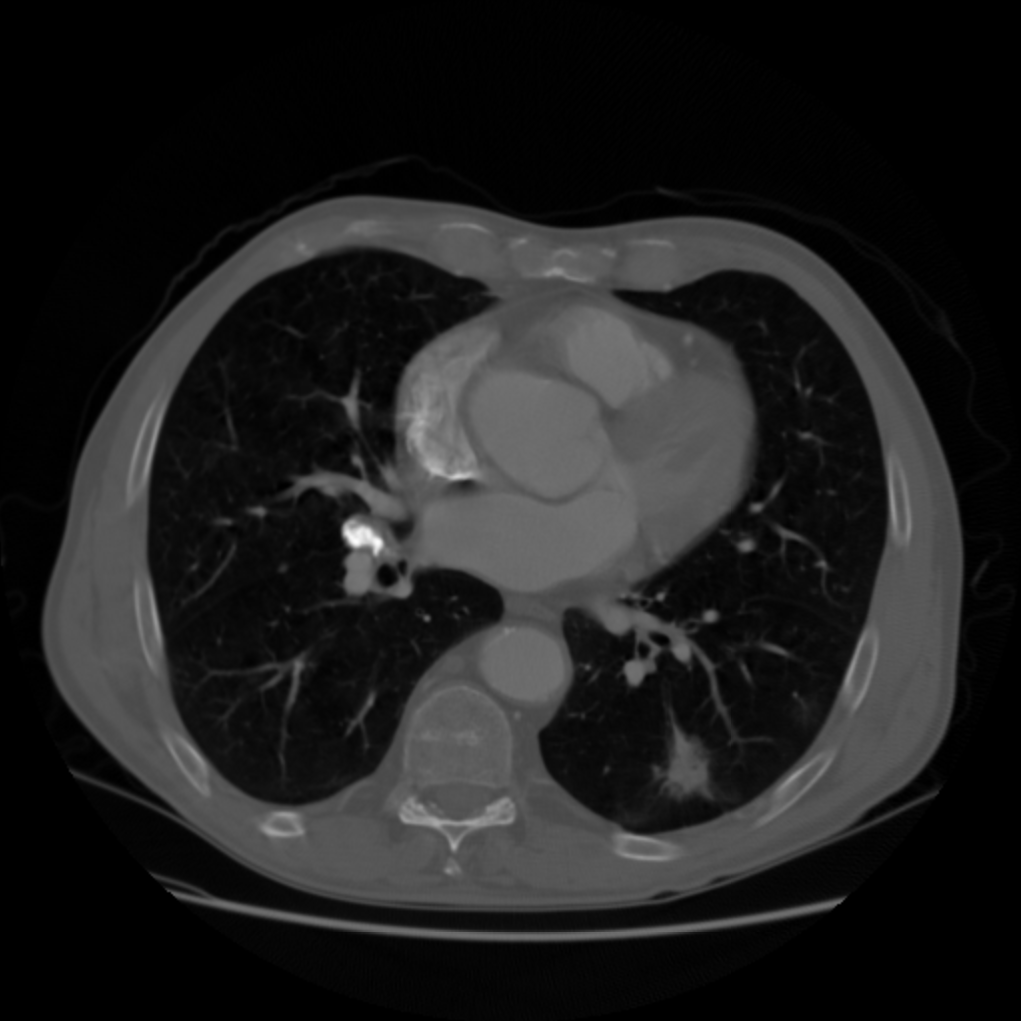}}
  \centerline{(c)}\medskip
\end{minipage}
 \vspace{-0.5cm}
\caption{The three images used in our experiments. (a) Spots. (b) Forbild phantom. (c) A real CT image.}
\label{Phantom}
\end{figure}

We compare the computation time and SNR of back projection among oblique method, orthogonal method under band limited assumption and box spline methods. We throughly test the performance of oblique method, box spline methods and blurred box spline method (box spline method with detector blur effect). Three images are used in our experiments, the first one consists of randomly placed ellipses, the second one is Forbild head phantom, and the last one is a real CT image from LIDC-IDRI dataset \cite{article} (Fig.\ref{Phantom}). In our experiments, we fix the sampling steps to be equal in all dimensions, $\boldsymbol{\Lambda}_x = {\lambda}_{x}I_2$ and $\lambda_y = n \lambda_x$, where $n$ is the downsampling rate. The number of views is set to $180$. Signal to noise ratio (SNR) in dB and structural similarity index (SSIM) are measured for all the methods. All computations were done on one PC (Windows 10 with a 3.7GHz 6-Core Intel Core i7-8700K processors and 32GB of RAM) using Matlab.

\vspace{-0.5cm}
\subsection{Speed and Accuracy of Back Projection}
To test the speed of different methods, we compute back projections with reconstruction size of $64^2, 128^2, ..., 1024^2$ pixels, and the downsampling rate $n$ is set to $1$. To test the accuracy of these methods, we set the reconstruction size to $64 \times 64$ and use a $640 \times 640$ image to calculate the ground truth. We vary the downsampling rate between $0.5$ and $2$.

\begin{figure}[htb]

\begin{minipage}[b]{.48\linewidth}
  \centering
  \centerline{\includegraphics[width=4.6cm]{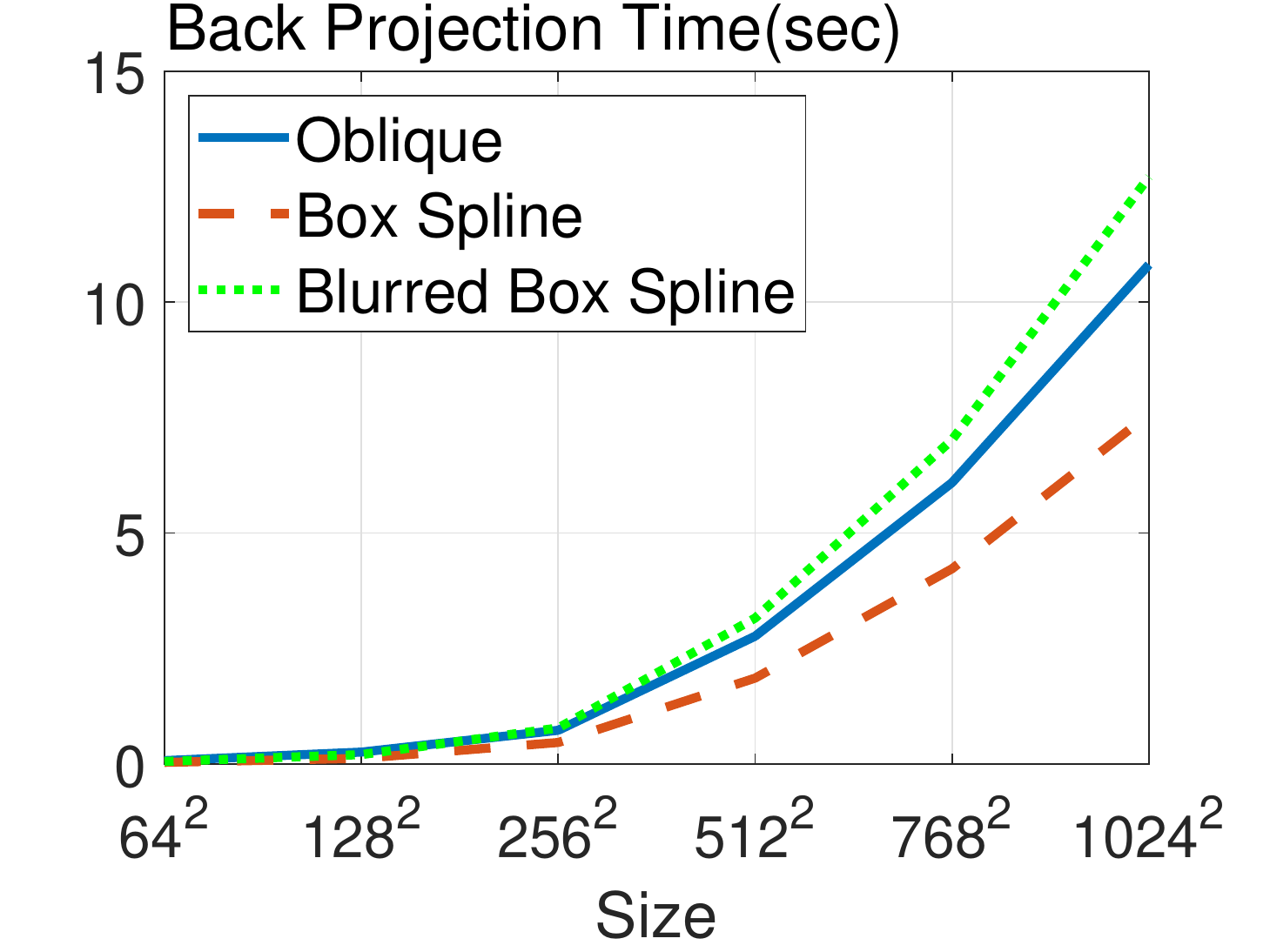}}
  
\end{minipage}
\hfill
\begin{minipage}[b]{0.48\linewidth}
  \centering
  \centerline{\includegraphics[width=4.6cm]{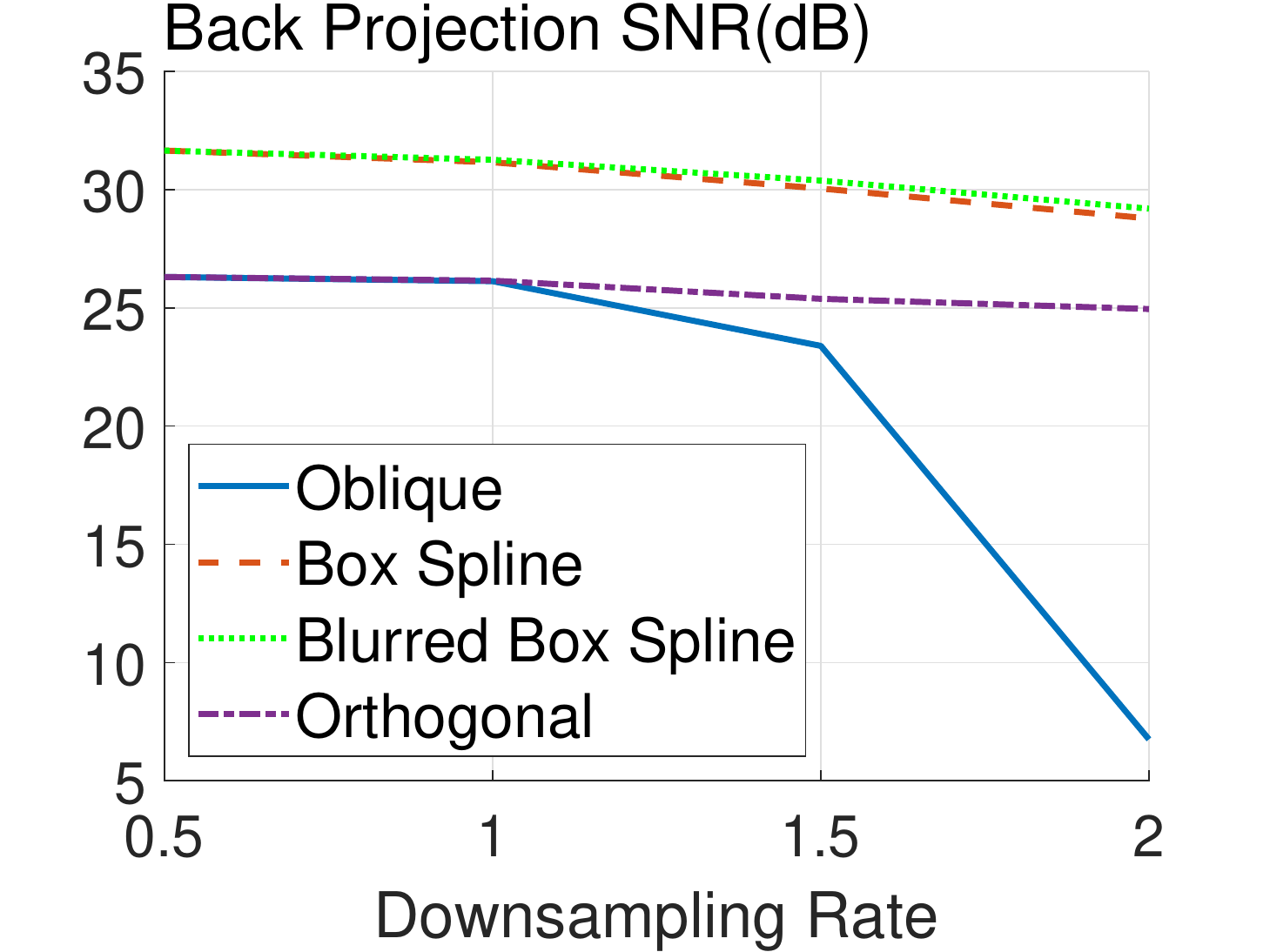}}
  
\end{minipage}
 \vspace{-0.5cm}
\caption{Computation time and accuracy comparison of back projection. The computation time of orthogonal method is too long to be plotted.}
\label{backproj}
\end{figure}

Fig.\ref{backproj}a shows that box spline method provides an improvement in computational performance (at least $30\%$) compared to the oblique interpolation method. These methods do not model the detector blur and our blurred box spline model, which provides a more realistic forward model, comes with a slight increase in computational cost (about $20\%$ slower). This is in line with our analysis in Section \ref{sec:back projection} that the detector blur effect just increases the number of box spline directions by one and has no substantial effect on speed. The computation time of orthogonal method is too long for a practical method.

Fig.\ref{backproj}b shows that the SNR of box spline methods are about 5dB's higher than that of the methods under band limited assumption. Box spline methods perform much better than oblique method when downsampling rate is high. These are in line with our analysis in Section \ref{sec:back projection} that box spline methods utilize the fact that $g_\psi$ is composed of univariate box splines to increase accuracy, while other methods make band limited assumption thus decreasing accuracy.
\vspace{-0.25cm}
\subsection{2D Reconstruction}
To test accuracy of reconstruction, we set reconstruction size from $64 \times 64$ to $1024 \times 1024$ and downsampling rate from $0.25$ to $2$. We compute unregularized reconstructions by using steepest descent algorithm \cite{Shewchuk94anintroduction}. We calculate the SNR and SSIM between the reconstruction $\hat{f}_\psi$ and the ground truth $f_c$. The ground truth is discretized by pixel-basis with ten times the resolution of the reconstruction and the X-ray transform is calculated exactly so that the inverse crime is avoided.

\begin{figure}[htb]
\centering
\hfill
\begin{minipage}[b]{.3\linewidth}
  \centering
  \centerline{\includegraphics[width=3.1cm]{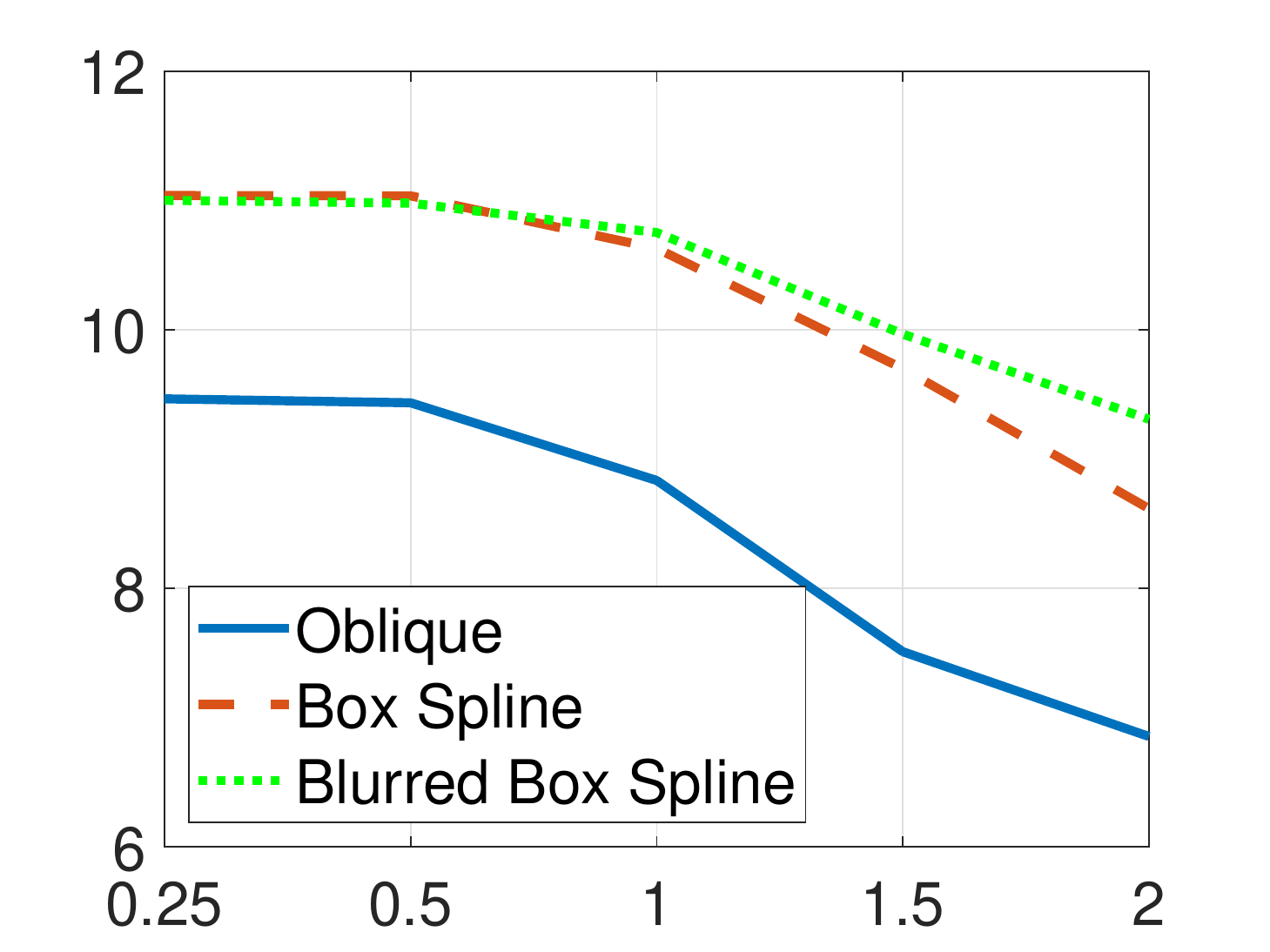}}
\end{minipage}
\hfill
\begin{minipage}[b]{.3\linewidth}
  \centering
  \centerline{\includegraphics[width=3.1cm]{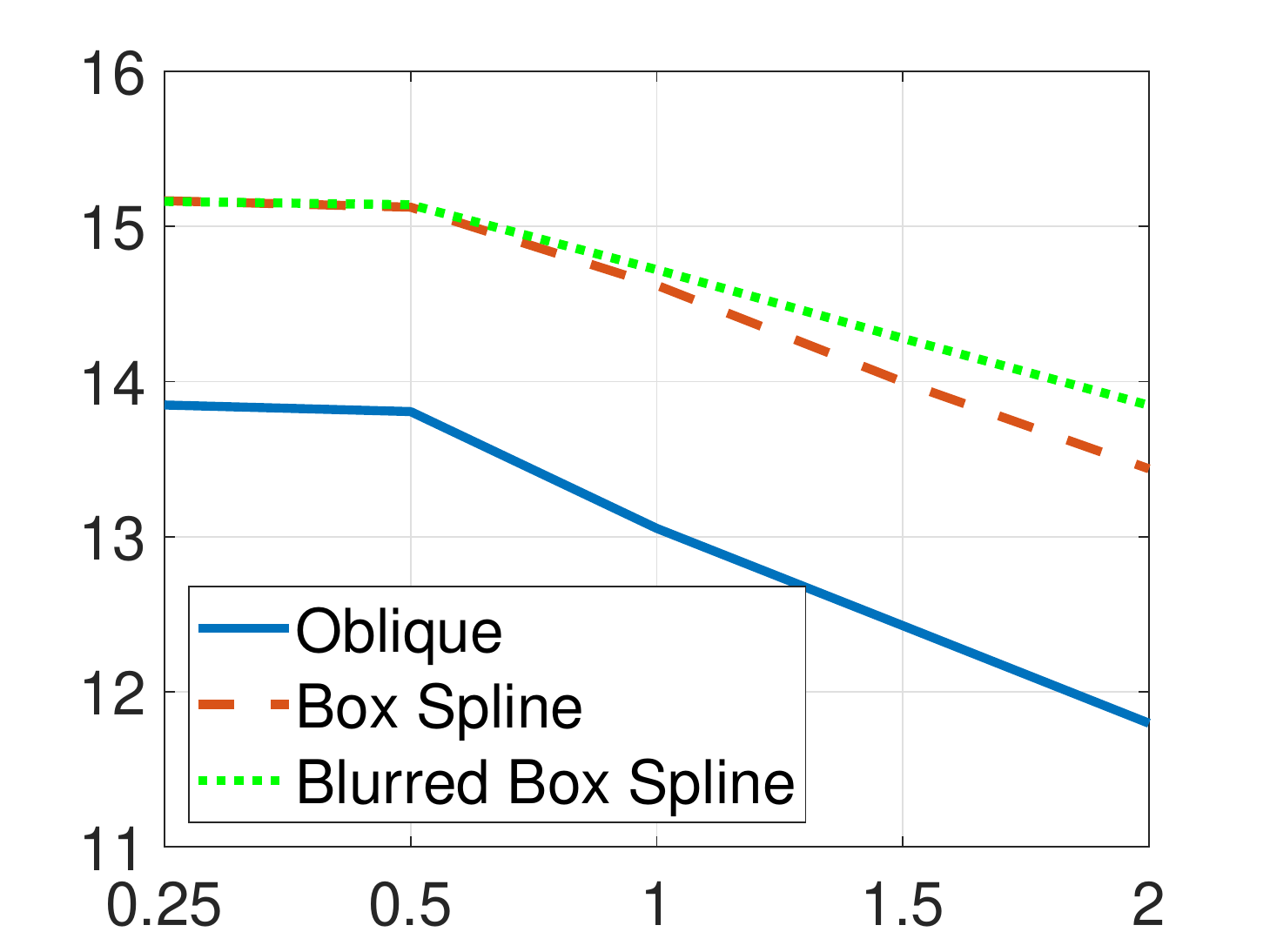}}
\end{minipage}
\hfill
\begin{minipage}[b]{0.3\linewidth}
  \centering
  \centerline{\includegraphics[width=3.1cm]{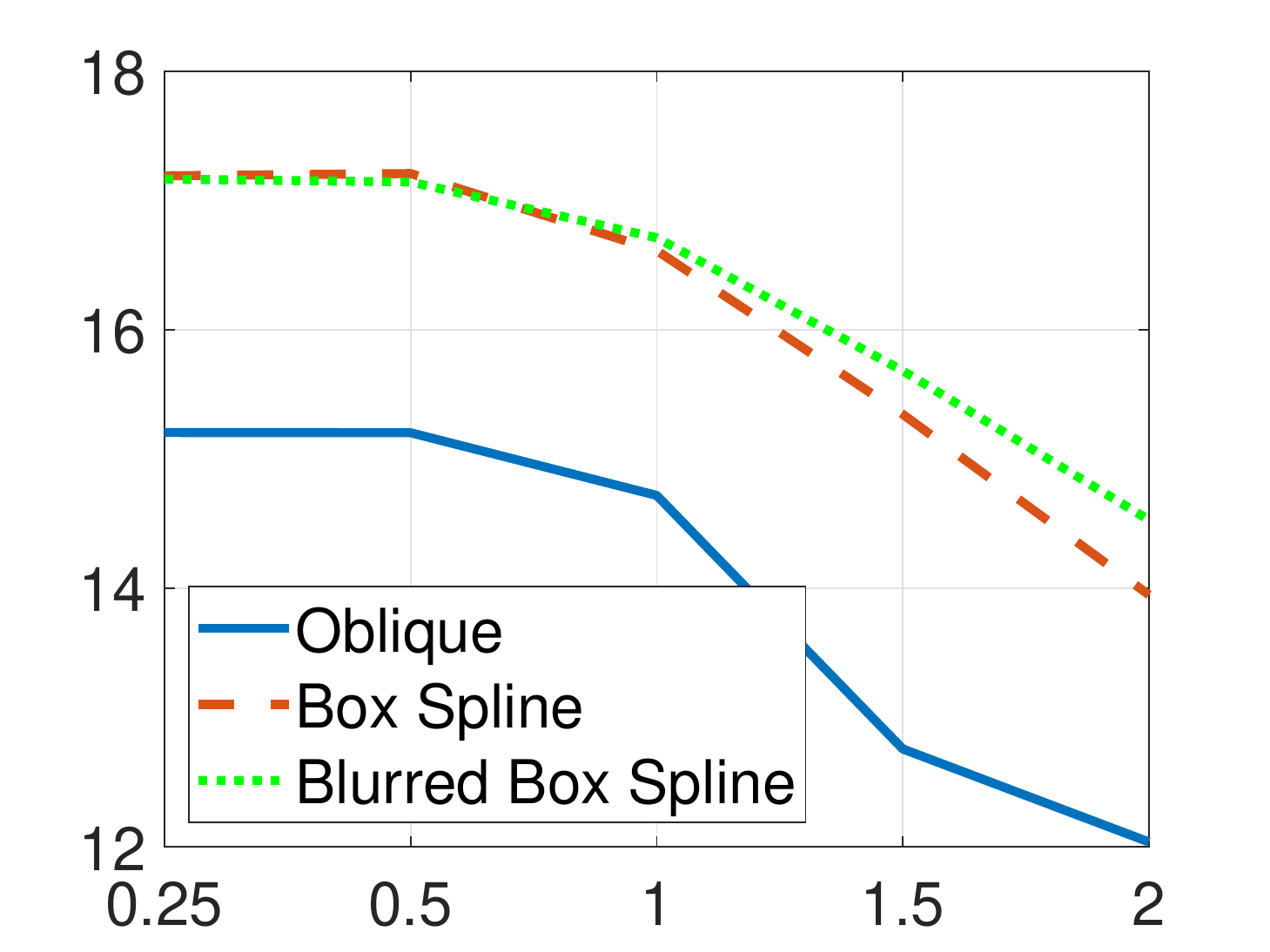}}
\end{minipage}

\hspace{0.08cm}\hfill
\begin{minipage}[b]{.3\linewidth}
  \centering
  \centerline{\includegraphics[width=3.1cm]{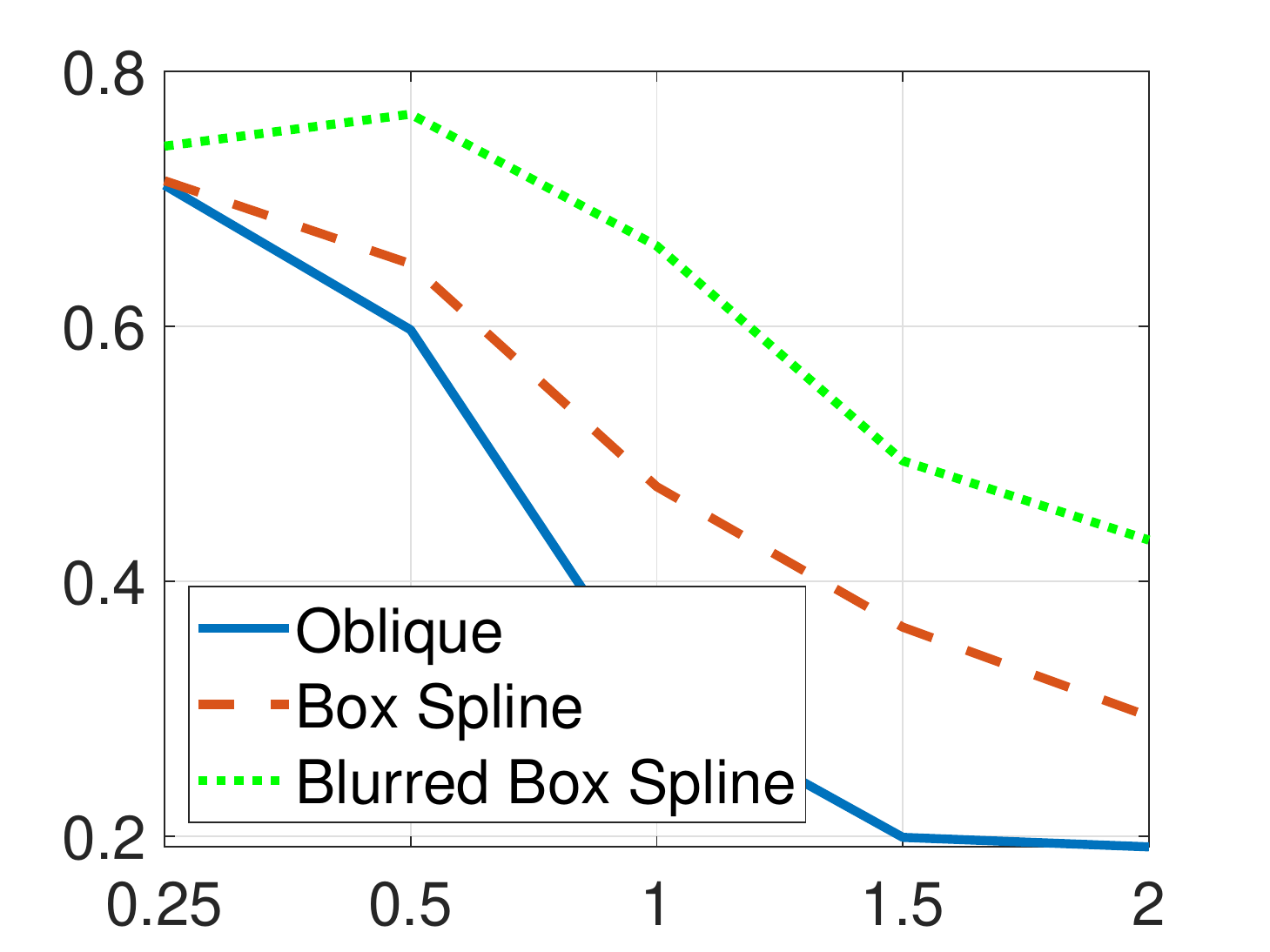}}
  \centerline{(a)}\medskip
\end{minipage}
\hfill
\begin{minipage}[b]{.3\linewidth}
  \centering
  \centerline{\includegraphics[width=3.1cm]{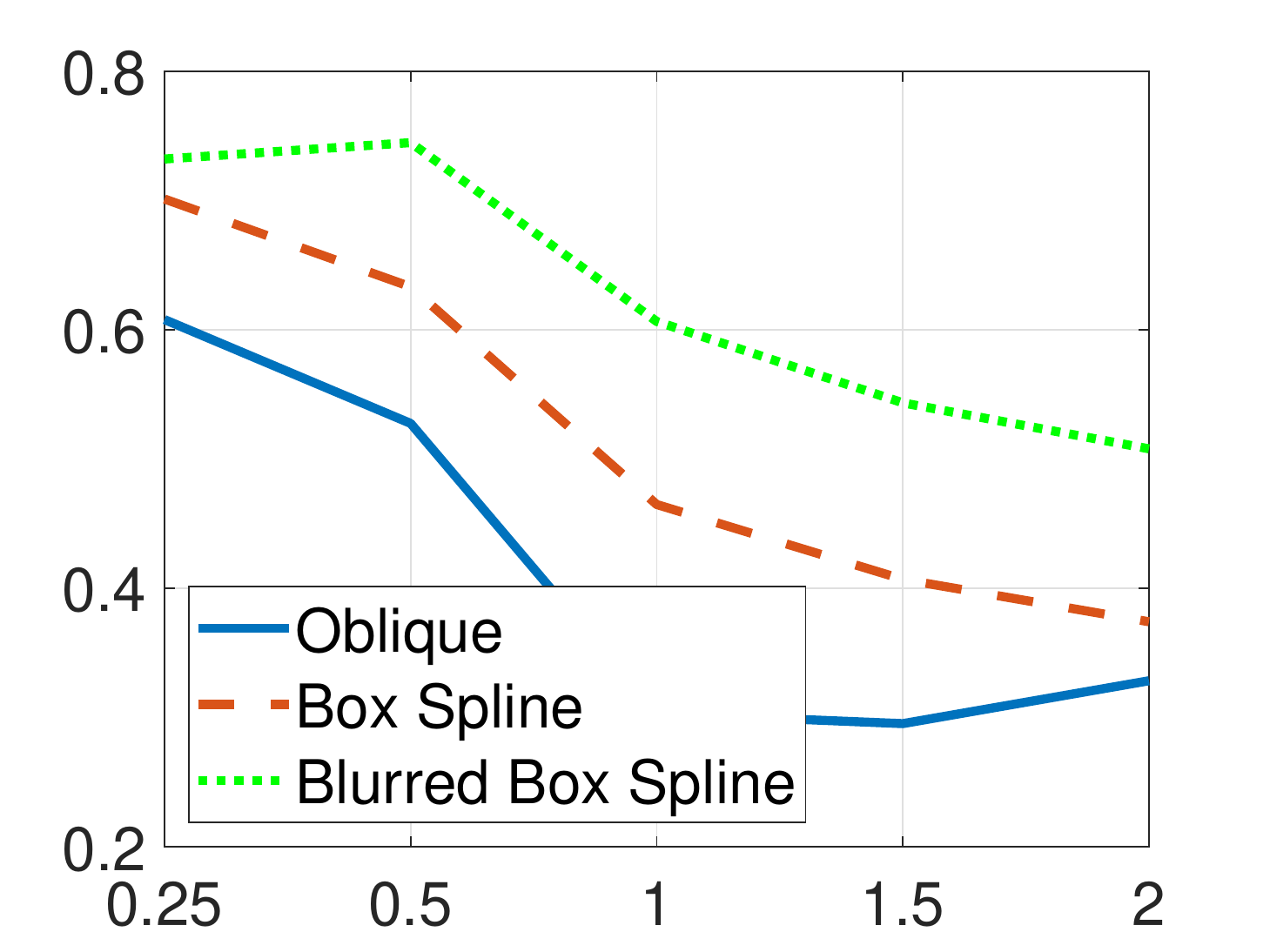}}
  \centerline{(b)}\medskip
\end{minipage}
\hfill
\begin{minipage}[b]{0.3\linewidth}
  \centering
  \centerline{\includegraphics[width=3.1cm]{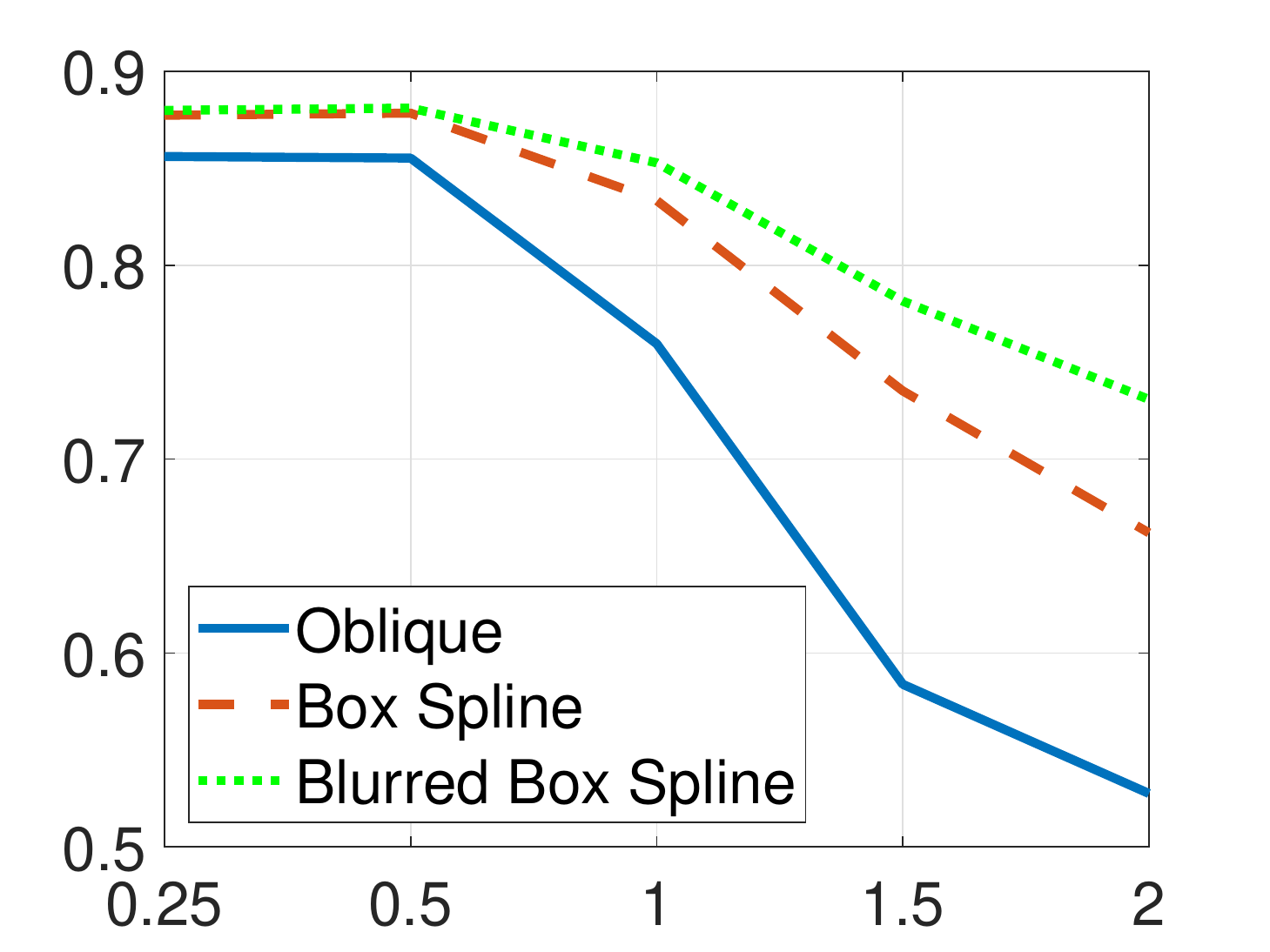}}
  \centerline{(c)}\medskip
\end{minipage}
 \vspace{-0.5cm}
\caption{Effect of downsampling rates on the accuracy of reconstruction. The first row shows SNR(dB) and the second row shows SSIM. (a) Spots. (b) Forbild phantom. (c) A real CT image.}
\label{recon vs downsampling rates acc}
\end{figure}

\begin{figure}[htb]

\begin{minipage}[b]{.48\linewidth}
  \centering
  \centerline{\includegraphics[width=4.5cm]{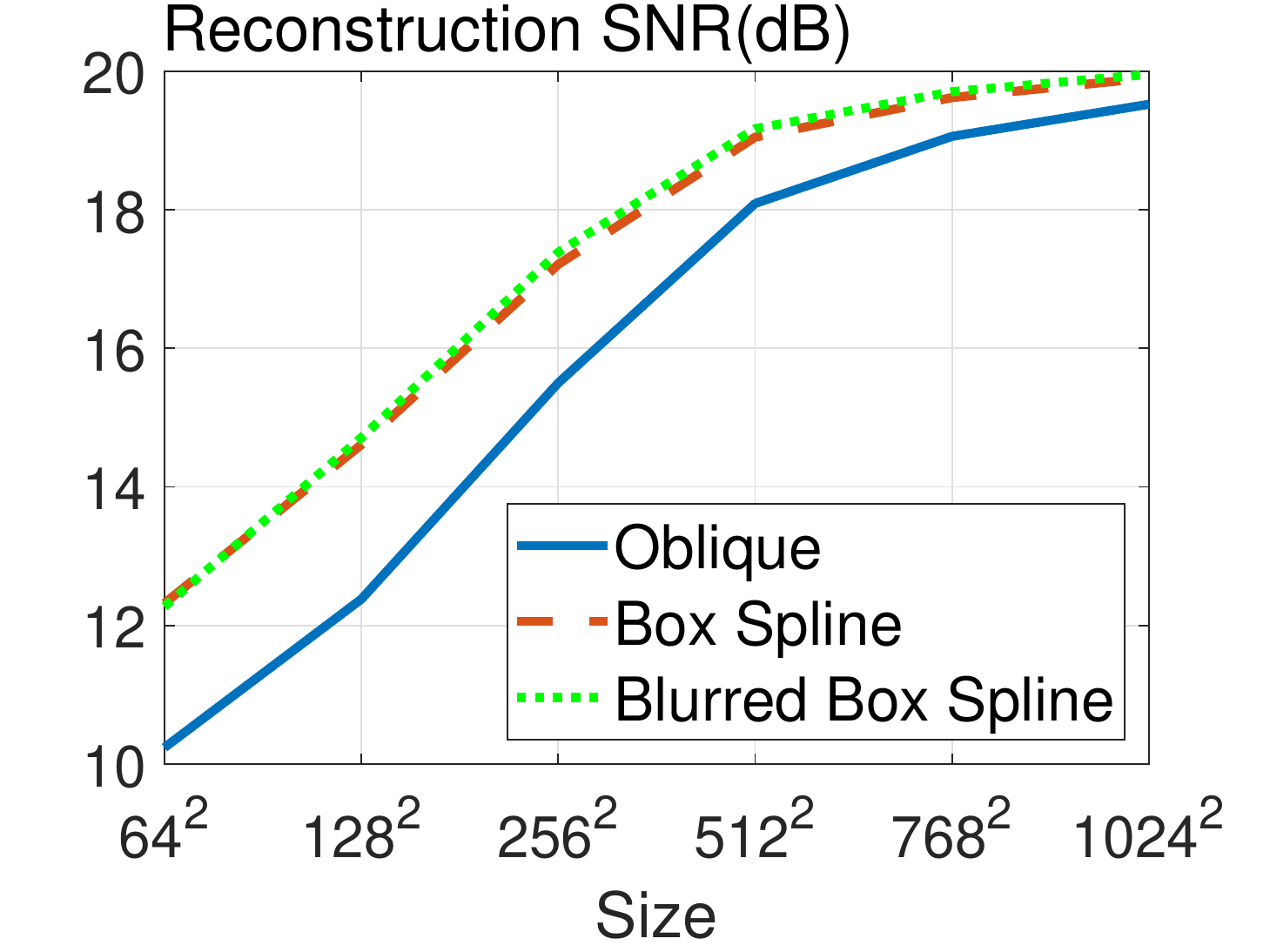}}
\end{minipage}
\hfill
\begin{minipage}[b]{0.48\linewidth}
  \centering
  \centerline{\includegraphics[width=4.5cm]{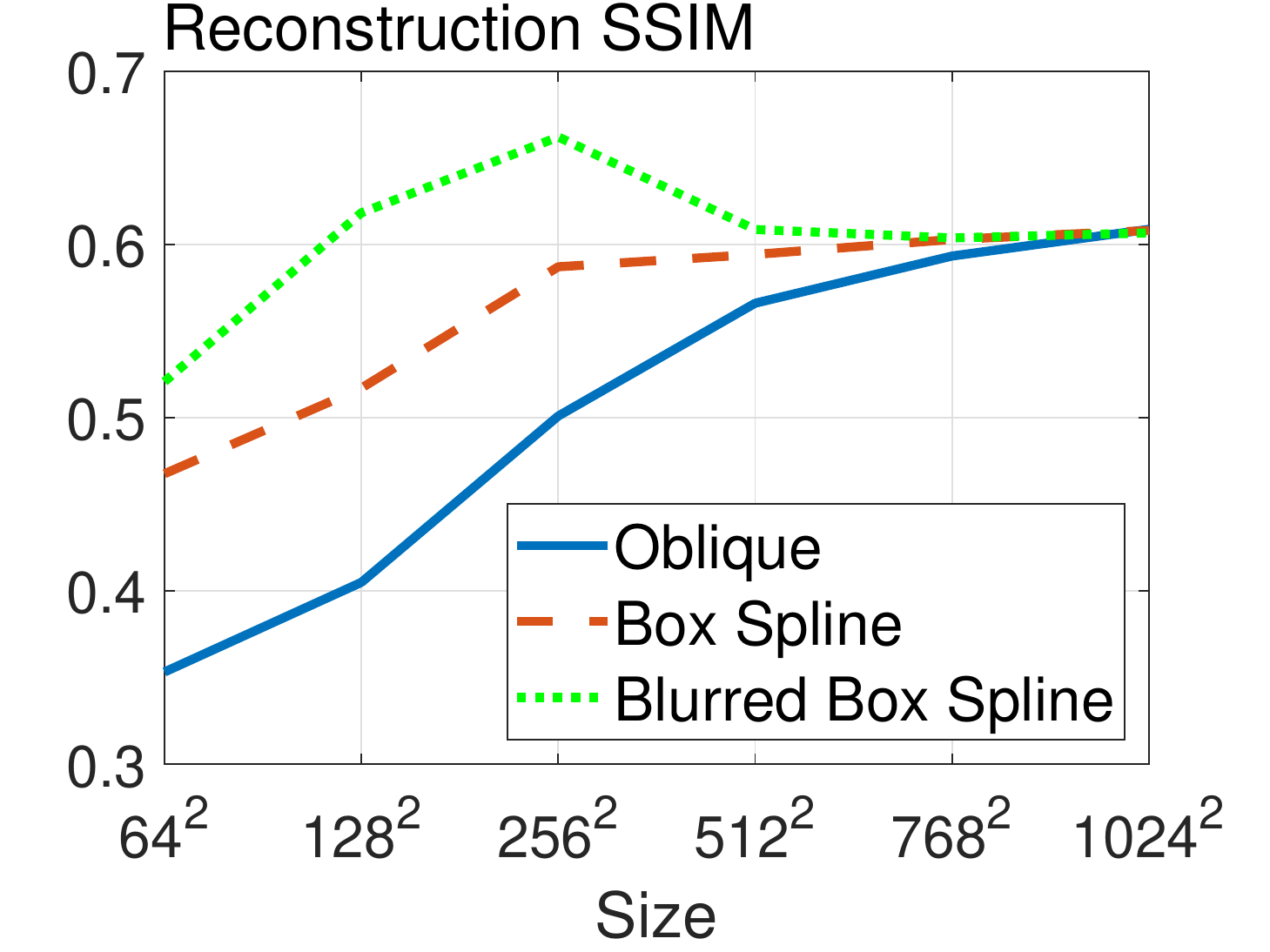}}
\end{minipage}
 \vspace{-0.5cm}
\caption{Reconstruction SNR and SSIM comparison.}
\label{recon vs size}
\end{figure}

\begin{figure}[htb]
\begin{minipage}[b]{0.3\linewidth}
	\centering
	\centerline{\includegraphics[width=2.7cm]{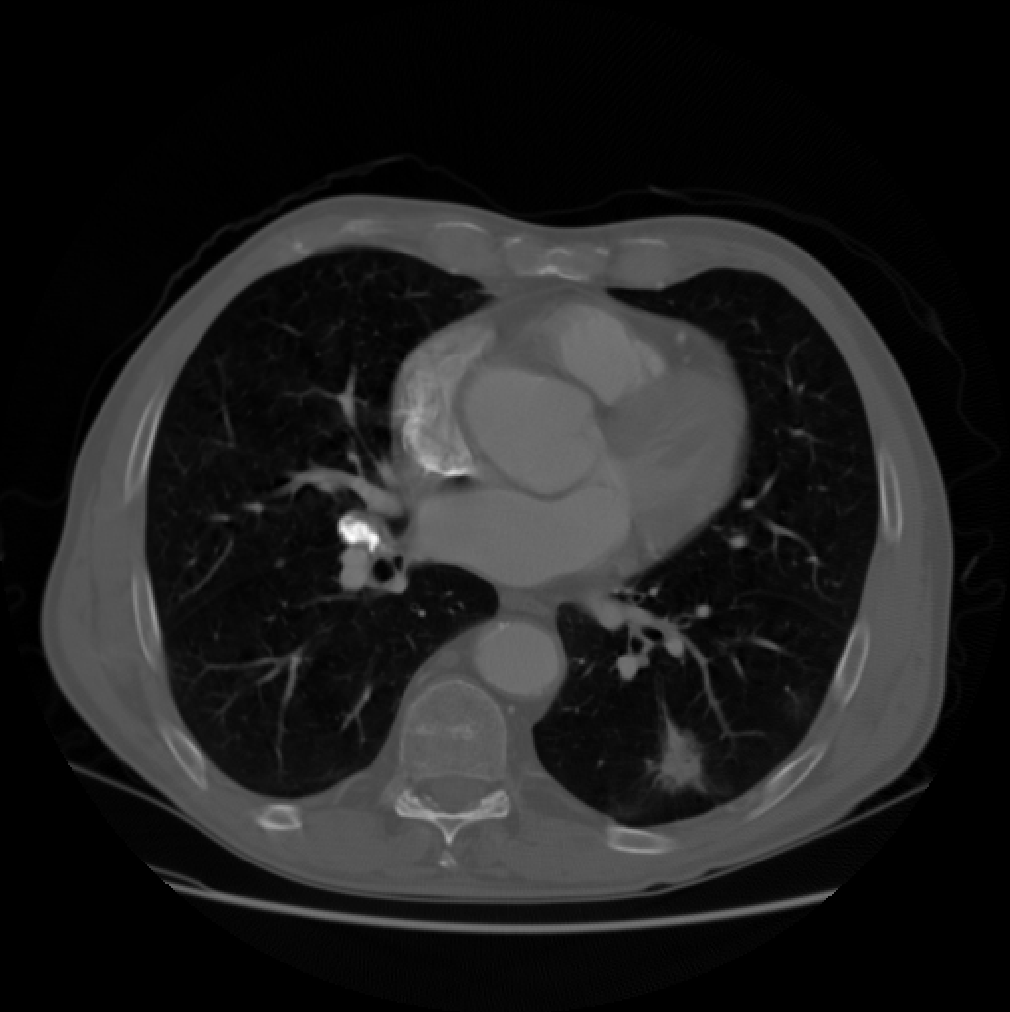}}
	\centerline{(a)}\medskip
\end{minipage}
\begin{minipage}[b]{0.3\linewidth}
  \centering
  \centerline{\includegraphics[width=2.7cm]{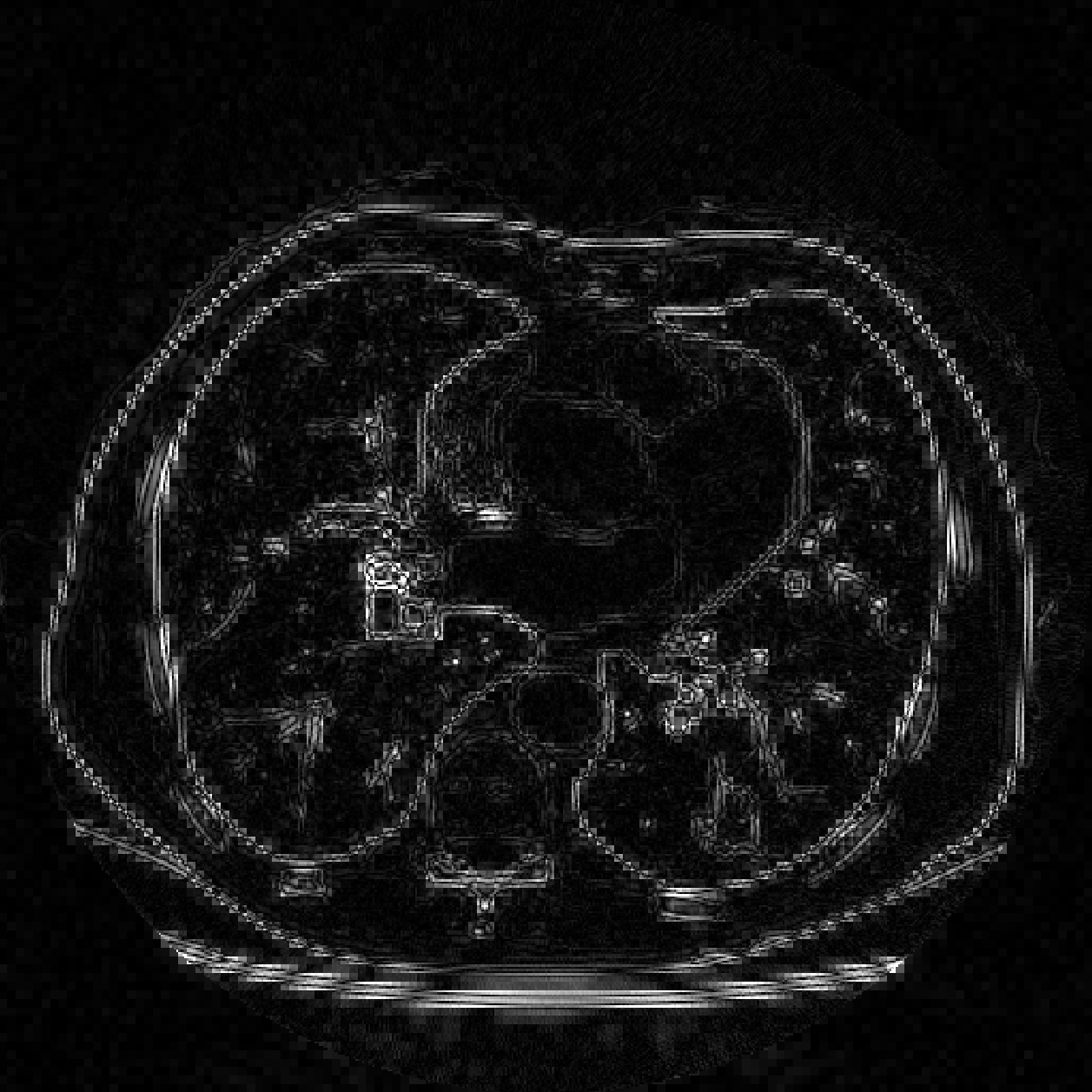}}
  \centerline{(b)}\medskip
\end{minipage}
\begin{minipage}[b]{.3\linewidth}
  \centering
  \centerline{\includegraphics[width=2.7cm]{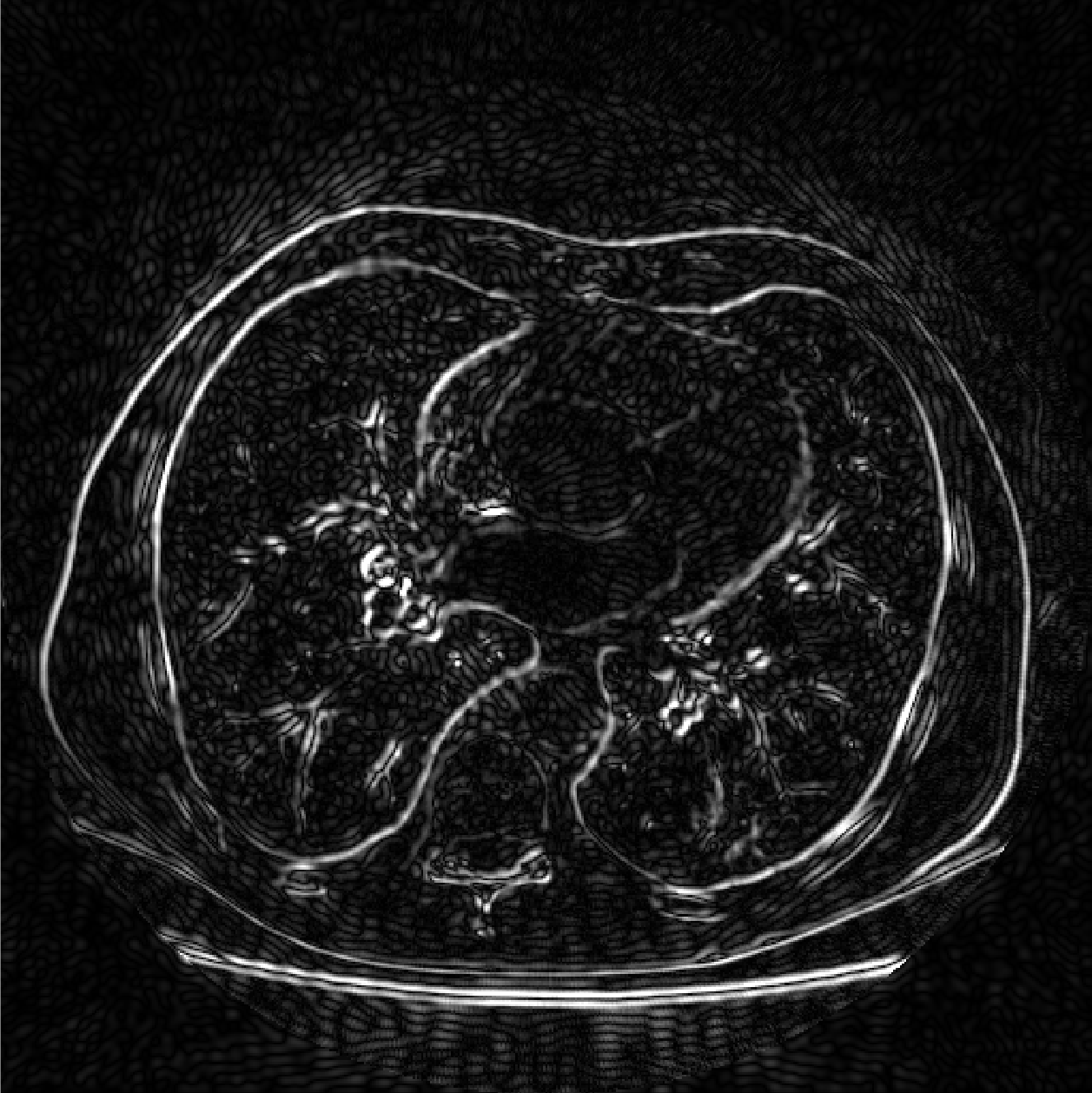}}
  \centerline{(c)}\medskip
\end{minipage}
 \vspace{-0.5cm}
\caption{Absolute total error $E_{\rm total}$ for the oblique method and box spline method on a real CT image. (b) and (c) are shown in the same scale to show the relative size of errors. (a) Ground truth. (b) Box spline, SNR = 16.7dB, SSIM = 0.84. (c) Oblique, SNR = 14.7dB, SSIM = 0.75.}
\label{real_img}
\end{figure}

Fig.\ref{recon vs downsampling rates acc} shows that our methods provide an improvement over the oblique method in our test cases. Fig.\ref{recon vs size} shows that our methods are more consistent when the resolution is relatively low. The reason is that oblique method truncates and estimates the sinc function, while we calculate the box spline exactly and have more accurate back projection. Fig.\ref{real_img} shows the absolute total error $E_{\rm total}$ of the real CT image, the image reconstructed by oblique method suffers from higher error.
\vspace{-0.25cm}
\subsection{Effect of Noise}
\begin{figure}[htb]
\vspace{-0.5cm}
	\begin{minipage}[b]{1\linewidth}
		\centering
		\centerline{\includegraphics[width=7.0cm]{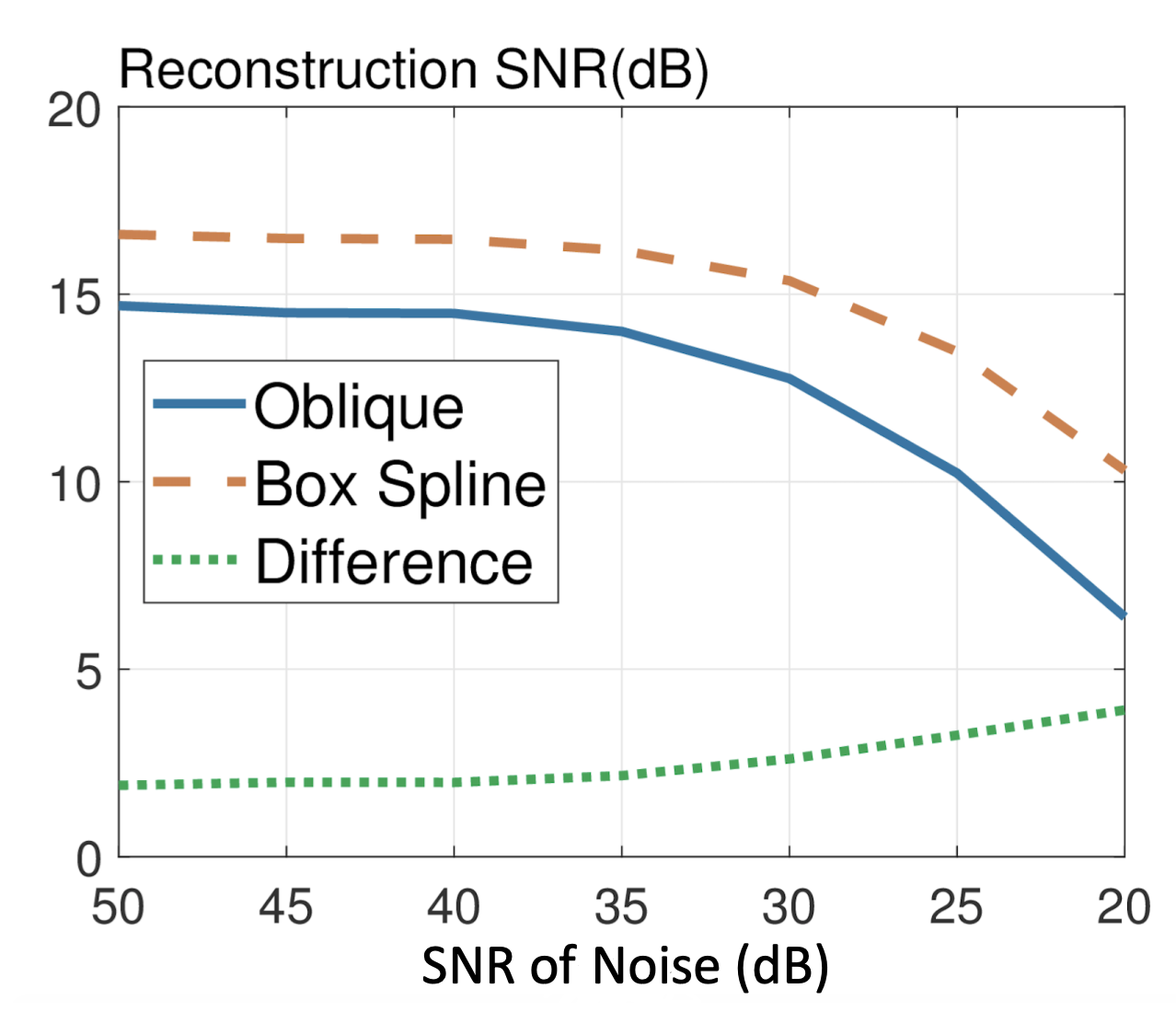}}
		  \vspace{-0.5cm}
	\end{minipage}
	\caption{Effect of sinogram noise on reconstruction accuracy.}
	\label{noisy recon}
\end{figure}

Now we repeat the 2D reconstruction experiment but with different levels of  Gaussian noise added to the sinogram. The result is shown in Fig.\ref{noisy recon}. It turns out the box spline method is less sensitive to noise.

\section{conclusions}
\label{sec:majhead}

Pixel-basis allows us to calculate the Gram filter exactly, choose the best interpolation kernel without any estimation, and model detector blur effect with high efficiency. In this paper, we compare our kernel with the sinc kernel, which provides the best discretization of band limited signals. The experiments show that using pixel-basis improves speed and accuracy of back projection and reconstruction, and this improvement is most evident with low discretization rates and sampling rates. Furthermore, we can model detector blur effect without losing speed and accuracy. With these results, we conclude that the proposed methodology improves the accuracy of the method based on the band limited assumption.

\newpage

\bibliographystyle{IEEEbib}
\bibliography{strings,refs}

\end{document}